\documentclass[3p,times,procedia]{elsarticle}
\usepackage{ecrc}
\usepackage{tikz} 
\tikzset{mynode/.style={font=\footnotesize,inner sep=0pt,text=black}
}
\usepackage{amsmath}

\volume{00}

\firstpage{1}

\journalname{Procedia Computer Science}

\runauth{V. De~Florio}

\usepackage{fancyhdr}
\jid{procs}

\usepackage{amssymb}

\usepackage[figuresright]{rotating}

\begin{document}

\begin{frontmatter}



\dochead{4th International Workshop on Computational Antifragility and Antifragile Engineering (ANTIFRAGILE 2017)}

\title{Systems, Resilience, and Organization:
Analogies and Points of Contact
with Hierarchy Theory}


\author[a,b,c]{Vincenzo De~Florio\corref{cor1}}

\address[a]{VITO, Boeretang 282, 2400 Mol, Belgium}
\address[b]{Global Brain Institute \& Evolution, Complexity and
COgnition group, Vrije Universiteit Brussel}
\address[c]{MOSAIC, University of Antwerp, 2020 Antwerpen, Belgium}

\begin{abstract}
Aim of this paper is
to provide preliminary elements for
discussion about the implications of the Hierarchy Theory of Evolution
on the design and evolution of artificial systems and socio-technical organizations.
In order to achieve this goal, a number of analogies are drawn
between the System of Leibniz; the socio-technical architecture known as
Fractal Social Organization; resilience and related disciplines; and Hierarchy Theory.
In so doing I intend to provide
elements for reflection and, hopefully, enrich the discussion on the above
topics with
considerations pertaining to related fields and disciplines, including computer science,
management science, cybernetics, social systems, and general systems theory.
\end{abstract}

\cortext[cor1]{Corresponding author. Tel.: +32-485-287633.}
\fntext[cor2]{}
\end{frontmatter}

\ead{vincenzo.deflorio@gmail.com}


\def\CU{\hbox{$\mathcal{C}\kern-.2ex\scriptstyle\cup$}}



\section{Introduction}
In this paper I draw a number of analogies between the System of 
Leibniz~\cite{leibniz2006shorter,Leilines,LeibnizLoemker,LeibnizParkinson,LeibnizCouturat,DF14c},
my work on Fractal Social Organization~\cite{DF13c,DFCBD12,DFSB13a,DFSB14},
resilience~\cite{DF14d,DF14a,DF14b,DF13b}, and
Hierarchy Theory~\cite{HT:TE14a,HT:Eld85,HT:Eld89,HT:Eld86}. Aim of this effort is to provide
elements for reflection and, hopefully, enrich the discussion on the above topics with
considerations pertaining to related fields and disciplines, including computer science,
management science, cybernetics, general systems theory and combinatorics.

The paper is structured as follows:
in Sect.~\ref{s:Leibniz} my focus is the general systems theory
of Leibniz. 
Fractal social organization
is the subject of Sect.~\ref{s:fso}.
Section~\ref{s:res} draws analogies between resilience, antifragility, and related concepts
and the Hierarchy Theory of evolution.
Conclusions are finally stated in Sect.~\ref{s:end}.

\section{Analogies with the System of Leibniz}\label{s:Leibniz}
Building on top of the Aristotelian concept of entelechy,
Leibniz introduces a gestalt~\cite{DF14c,Bou56} called \emph{monad}. The treatise of the monad
translates into a series of dualisms: a monad is
(i) a unitary and indivisible
``whole''; but it is also a network of constituent ancillary parts;
a metaphysical
conceptual entity, which admits multiple physical
manifestations
(``encodings'');
a static, incorruptible model,
residing in a limitless, unconstrained,
time-and-space-less context; at the same time,
it is also an object reified and deployed as a dynamic entity
into a time-and-space
resource-limited ``world'' subjected to: constraints; a limited span;
strict energy requirements; and a corruptible structure.
Due to the above duality, a monad is characterized by 
two quality components:
(1) A static (immutable) quality component, representing the \emph{intrinsic quality\/}
of the model.
(2)item A dynamic and context-dependent quality component, representing
the extrinsic, or \emph{contingent quality\/} of the physical representations of the
model when set to operate in a given ``world'' or environment.
The first component corresponds to the systemic class of the model according to
some General Systems Theory classification---for instance
the behavioral classification in~\cite{RWB43} or the 
behavioral/organizational classification of~\cite{Bou56}.
The second component corresponds to a monad's instance's ability to match
the mutating constraints and circumstances expressed by the environments that
instance is set to operate in.

Quality is an \emph{essential\/} aspect in the Leibnitian System, in that
quality determines the existence of monads. In fact, according to Leibniz,
despite their residing in a meta-physical, timeless domain, monads
may exist or cease to exist. The only one to decide on the fate of monads is
the supreme monad, ``God'', who is the monad
representing the concept of 
``all that is\footnote{%
	Interestingly enough, already in~\cite{GuruGranthSahib}
	a similar concept was expressed, albeit of course in a poetic form.
	Among the lines at p.~1156 we have, e.g.,
	``Such is my Lord and Master, the Lord of the Universe'';
	``Millions of universes are the limbs of His Being'';
	``His Wondrous Plays are enacted on millions of stages''; and
	``Millions of expanses are His; there is no other at all''.}''
and is the only entity able to assess
both quality components of other monads. If a monad is ascertained as being
sufficiently ``qualified'', it is persisted---it ``stays in the mind of God'';
otherwise, it is not retained and ceases to exist. ``Qualified'' here
refers to a monad's \emph{quality of emergence\/} (QoE), namely 
\emph{how much\/} the whole represented by that monad is greater than the
sum of its parts. As discussed, e.g., in~\cite{DF17QoE,DF14b}, QoE can not be
assessed exclusively considering intrinsic qualities, hence a trial is necessary.
The worth of the monads must be tested by observing the sustained action of their
instances when deployed in a finite and mutating
environment. As many monads as possible will be
retained, but no more than it is possible given the constraints and limitation of the world,
also considering social aspects such as the possibility for coexistence (cf. tropism). 
``Compossibility'' is the term
used by Leibniz to refer to possibility for coexistence.


The Leibnitian God is in fact
the entity responsible for sorting out the monads worth
persisting and discarding the ``unworthy'' ones---that is, for the realization
of this ante-litteram ``survival of the fittest'' strategy~\cite{DF16L}.
In order to persist, the monad must pass through several
``sortings''. In fact, as suggested in~\cite{DF14c}, the role of God
in Leibniz may be described as that of a ``Universal Sort'' process---a
\emph{mechanism},
meant to evaluate each and every monad's two components of QoE
and to schedule for existence those monads that best-score, up to a
threshold 
expressed by the limitations of the current world. 
A pseudo-code for Ultimate Sort is available in~\cite{DF14c}.

Code is in fact another key ingredient in the Leibnitian System.
Code is the physical representation of a monad, and it also represents
a scheme for the physical construction of an instance---its realization,
or reification. Leibniz foresaw the existence of a universal ``language''
for the expression of monads.
He called such language
\emph{Characteristica Universalis\/} (\CU).
\CU{} is a diagrammatic language employing pictograms. The pictograms are convenient representations
of modular knowledge of any scale, with segments representing different properties---for
instance
compossibility or its opposite. Leibniz exemplified this through diagrams
such as the one on the
frontispiece of his \emph{De Arte Combinatorica}.

The \CU{} language is Leibniz's way to encode monads as networks of other substances,
together with their relationships.
Pictograms represent \emph{modules}, namely knowledge-components packaging other ancillary knowledge-components.
As observed in~\cite{DF14c}, pictograms are Leibniz's equivalent of: Lovelace's and Turing's tables of instructions;
subroutines in programming languages; boxes in a flowchart; or components in component-based software
engineering. They represent a \emph{hierarchy\/} of related concepts producing a
whole characterized by some degree of QoE~\cite{DF17QoE}.

A \CU{} ``code'' is the physical representation of a ``metaphysical'' concept---viz., of a
monad. It is the representation of a model, expressed in terms of relationships
with other models and in an abstract and static way, independent of whatever the
physical
``ambient'' and whatever the scale. 
Compositional and modular by construction, 
\CU{} is the
language of the 
``true characteristic [of the monads,]
which would express the composition of concepts by the combination
of signs representing their simple elements, such that the correspondence
between composite ideas and their symbols
would be natural and no longer conventional.''~\cite{LeibnizCouturat}

In other words, \CU{} is 
an isomorphic language, such that \emph{concepts are preserved through their compositions}.
Instances are thus ``code'' expressing a model: in other words, a phenotypical
representation of a genotype.

\subsection{Analogies with Hierarchy Theory}

From the above discussion one can clearly see that the duality expressed
in the System of Leibniz may be put
into direct correspondence with the ``double hierarchy'' 
at the core of Hierarchy Theory (HT).
In HT, two concurrent and intertwined hierarchies define the action of nature at all
scales: a ``genealogical hierarchy'' (GH), responsible for the trustworthy transmission
of hereditary characters through time and a so-called ``economic hierarchy'' (EH),
responsible
for the physical construction of individual phenotypical instances of a genetic ``model''
as well as for guaranteeing the premises for a trustworthy transmission of
the hereditary traits.

\subsubsection{Genealogical Hierarchy}
I conjecture that the purpose of GH may be put in relation with the Leibnitian 
concept of the intrinsic quality of a monad. 
As already mentioned, monads in Leibniz are pure, immaterial concepts;
though this is what Leibniz refers to
as ``reality\footnote{%
	Again this concept is already present, e.g., in Sikhism~\cite{GuruGranthSahib}. As an example,
	the term ``avatar'' stands for ``a deliberate descent of a soul to earth in any form''.\label{fn:avatar}}''.
But in order to exist, reality needs some form of materialization.
In other words, a monad needs to be represented in a physical way---by means
of a physical substance that
encodes and ``embodies'' it. Putting it in a different way,
in order for the class to exist,
there must be a method to produce individual resilient instances of that class.
Those instances constitute the \emph{essence\/} of their class, and
vice-versa---one cannot exist without the other.
Instances become
``identifiers'' (or avatars; see Footnote~\ref{fn:avatar}) of their class.

This concept is nicely rendered
in Algebra: given any non-empty set $S$ and any partition of $S$ defined by an
equivalence relation $R$, for any instance $x\in S$ the equivalence
class of $x$ (namely the block of $S$ $x$ belongs to) is
simply $[x]_R$: any element $x$ is representative of the class it belongs to.
The same conclusions may be reached considering that the definition of
the projection function $\pi: S\rightarrow S/R$ is simply
\( \forall x\in S: \pi(x) = [x]_R. \)

With the terminology of
Hierarchy Theory, instances are \textbf{replicators} of their class.
They must be resilient, because
their demise would translate in the extinction of their class---their monad.
The problem is then being able to persist the concept throughout time, while ``moving''
through a medium that affects the instances in several ways. 
This is one aspect of \emph{resilience}, viz. ``being at work while staying the same''.
The term ``replicator'' implies the only effective strategy for a monad
to ``stay the same'':
redundant copies of the model of the whole and of its constituent parts must be created
and propagated through time by means of an
uninterrupted ``chain'' of replicas---think for instance of
the incessant work of amanuensis monks.
Of course in order to be effective, the strategy requires that each new copy,
or \emph{offspring}, be ``compliant'' to its parent.
In mathematical terms, given a function \CU{}, we can model a genealogical
evolution of replicator instances as the orbits of a dynamical system, namely
the recursive application of \CU; and the effectiveness of the strategy
corresponds to asking that, at any time $t$, the corresponding orbit of \CU{}
be compliant to that of its predecessor. I plan to develop this in more detail
as I did for the finite state automata and dynamical systems 
introduced in~\cite{DeFl96} and~\cite{DeFl05}.

Another way to discuss the effectiveness of the genealogical
strategy is given by introducing the concept of \emph{fidelity},
which we defined in~\cite{DF14a,DBLP:journals/corr/FlorioP15} as the ``compliance between corresponding
figures of interest in two separate but communicating domains''.
Replicators are effective if, at any given time, there exists at least
one replicator and that replicator is characterized by fidelity. In other words,
the first-generation copies must be
faithful representations of the model, the second-generation copies must
be faithful representations of the first-generation copies, and so forth.
A transitive closure of isomorphic transformations must be valid
across all generations in order to guarantee that
at any given time $t$ the ``copy-at-$t$'' instance
faithfully encodes the original model---or,
in other words, that it represents the same concept.
Only when this transitive closure holds we can guarantee the transmission
of the monad through time\footnote{%
	The concept of replicator has been exemplified in popular
	literature by characters such as
	``the Phantom'' by Lee Falk~\cite{Phantom}. The Phantom is a man that embodies
	certain values---fighting injustice, greed, and violence. In order to guarantee
	the persistence of this ideal, the Phantom is a replicator: a genealogy of Phantoms
	exists, the current Phantom being the 21\textsuperscript{st} in a line that goes back to 1536.
	Since then, each Phantom is faithful to an oath made by his predecessor, up till
	the original one who defined the oath and the genealogical strategy. As long as 
	a living offspring exists and the
	``genetic oath'' is faithfully renewed, the Phantom concept is persisted. As a result of this,
	people unaware of the origin of the Phantom consider it as 
	``\emph{the man who cannot die}.''}.

Finally, GH can provide a convenient,
``Pandaeans'' interpretation of the Leibnitian concept of persistence-in-the-mind-of-God.

\subsubsection{Economic Hierarchy}
As a second conjecture, we relate the Leibnitian
concept of extrinsic (or contingent) quality
with the Hierarchy Theory concept of EH. 
Leibniz considers matter as ``a substance's privative or passive
aspects''~\cite{leibniz2006shorter}. In fact, it is the moment a model 
is real-ized (material-ized), that it needs
to confront oneself with physical limitations, deployment factors, design constraints,
social interaction, and other factors.
Furthermore, materialization means deployment in a multi-user environment whose
cohabitants all compete for the same objective: maximize their chances for survival
and for the persistence of their identity.
Leibniz explains that, for any given ``world'', places and energy are finite and
represent resources to compete for.
Those hard limitations call for economic considerations: redundancy, for instance,
cannot be unlimited, as each replica is associated with a ``cost''.
Being a system-of-systems, any new individual is an \emph{economic event\/}
that reverberates in all the hierarchical levels the constituent sub-systems
``reside'' in. And every deployment choice at every level translates
into the introduction of different sorting criteria---different
weaknesses and strengths, that is, with respect to the 
varying environmental conditions.

Leibniz does not introduce explicitly the HR concept of
\textbf{interactors}---``materializators'' that interpret the model building
instances and introduce physical differentiating factors; but he does
recognize that a code does not exist \emph{per se},
unless a reference ``machine'' exists
and is able to interpret and translate it into dynamic behaviors or into some other,
diverse but conceptually equivalent, form.
Therefore Leibniz introduces the concept of an interpreter for \CU{} codes.
It is the so-called Calculus Ratiocinator (CR), the reference ``hardware''
for the \CU{} ``language''. CR is also the algebra or ``validation environment'' where
instances are put to test and QoE is ascertained~\cite{LeibnizCouturat}.


\subsubsection{Emergence and Modularity in Leibniz}
\begin{figure}
	\centerline{\includegraphics[width=0.15\textwidth]{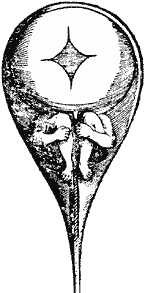}}
\caption{Preformationism exemplified by Nicolaas Hartsoecker, 1695. Image from the Wikimedia Commons.}\label{f:pref}
\end{figure}

As already mentioned, a monad is persisted in the ``mind-of-God''
only if it proves to be characterized by an adequate degree of
intrinsic (systemic) quality and extrinsic (contingent)
quality across a dynamic variety
of conditions~\cite{DF17QoE}.
Monads and their material instances across the global biota
are continuously ``sorted''---which constitutes for Leibniz
one of the purposes of ``God''. Said purpose is not only
effective but also ``necessary'', and results in ours being
``the best of all possible worlds''~\cite{L:Theodicy}.
I believe that the treatise on Hierarchy Theory provides elements to better
understand the ``necessity'' in the processes exercised
by the Leibnitian ``God'':

\begin{itemize}
\item First, ``God'' aims at guaranteeing that ``the greatest amount of essence
or possibility is brought into existence''. HT provides us with a
reason for this: the greatest
amount of essence or possibility corresponds to
the \emph{greatest area of morphospace}~\cite{HT:TE14a}. The greater such area, the more covered
is the space of all possible events that may affect the hierarchies of the biota.
Uncovered areas
may in fact correspond to the most appropriate natural
``configurations'' with respect to an
unprecedented or very rare environmental condition. In other words, seeking
the greatest amount of essence or possibility aims at increasing
as much as possible
the amount of diversity and disparity in all levels of the natural hierarchies,
diversity and disparity being the most effective ``lines of defense''
against events that may
affect simultaneously a large amount of individual instances.
The Permian-Triassic extinction event (P-Tr) is an extreme case of 
diversity/disparity failure. The widespread adoption of a same
``deployment solution''---the use of a mineralized skeleton---resulted in a 
common trigger ultimately producing one of the most devastating and widespread
correlated failures
of recorded natural history~\cite{Knoll,Eraclios13.11.12}.

\item Secondly, Leibniz ``hints'' at the fact that the process enacted by ``God''
results in the long run in greater and greater QoE:
%
``When the
tables of categories of our \emph{art of complication\/} have
been formed, \textbf{something greater will emerge}.''~\cite{LeibnizParkinson}

Nature's ability to construct ever more complex---ever more \emph{evolved}---entities
is thus another ``proof'' of both the effectiveness and necessity of God's behaviors.
It is possibly this the ``greater secret [that]
lies hidden in our understanding, of which these are but the
shadows'', which Leibniz referred to in~\cite{LeibnizLoemker}.

\item
The ability to develop ever greater QoE is possibly one of the reasons
behind the interest that Leibniz showed throughout his life
for the work of Thonis van~Leeuwenhoek. Leeuwenhoek had developed
a clever technique for the creation of microscope lenses. With his
microscopes he was the first to observe
microorganisms and spermatozoa; because of this he is now
generally considered as the father of microbiology.
Leeuwenhoek was also the grand developer of preformationism, the belief according to which
all beings are the development of preformed miniature-versions he called
``animalcules''. Animalcules are genotypical, first-order ``code'' producing
a phenotypical second-order ``code''---a more developed,
living instance of the animal monad.
Figure~\ref{f:pref} exemplifies preformationism showing an
animalcule (a so-called homunculus) within a spermatozoon.

Though obviously an incorrect and unscientific concept, preformationism contains
\emph{in nuce\/} the
principle of \emph{conservation of modularity}, viz. the property of conserving modularization when passing from
a genotypical representation (viz. a concept, i.e., an abstract and general template)
to a phenotypical representation
(namely a particular ``realization'', or concrete expansion,
of the template). This property, which may
be probably best represented through the mathematical concept of an isomorphism between genotypical and phenotypical
algebraic domains, is in fact compatible with the Leibnitian vision of substances as ``second-order scripts'' produced by
``first-order scripts''.
This conservation of modularity possibly hints at the reasons why evolution ``evolves'', and
why nature ``naturally'' develops ever more complex substances~\cite{WaAl1996}.
%

\end{itemize}

\section{Fractal Social Organizations}\label{s:fso}

A second field useful for reflections in the context of Hierarchy Theory
is in my opinion the bio-inspired
distributed organization that I called Fractal Social Organizations
(FSO)~\cite{DF13c,DFCBD12,DFSB13a,DBLP:journals/corr/FlorioSB14,DFSB14}.
FSO is a fractal organization: it is a hierarchy 
that ``is not based on the classic top-down
flow of control and bottom-up flow of feedbacks (autocracy)
but rather on a peer-to-peer approach where every node
in the hierarchy may play both management and subordinate
roles depending on the situation at hand (sociocracy)''~\cite{DFSB13a}
and a set of rules valid at each level of the hierarchy.
The peculiar aspect of FSO with respect to other fractal organizations
is indeed its set of rules---the so-called \emph{canon\/} of the
organization. The FSO canon states that, whenever an event occurs in a 
\emph{focal level}~\cite{HT:TE14a}, the event is resolved by identifying
roles to be assigned to a response protocol. At first roles are
sought in the focal level only (by means of
semantic service description and matching~\cite{DBLP:journals/corr/FlorioSB14,SDB13a}).
When roles cannot be appointed to holons in the focal level (that is, to
its nodes) a so-called
\emph{exception\/} takes place, meaning that a ``missing roles'' event
is propagated to the level immediately above the focal one. Because of this
\emph{upward causation}, the focality moves to this second level.
Again, roles are sought in the new focal level,
possibly leading to new exceptions and new
propagation of focality. This ``movement'' traces entities
from different hierarchical levels and enrols them into a temporary new
network, which I called ``Social Overlay Network'' (SON).
The SON is the team of cross-level nodes that are to deal with the
originating event (the event that occurred in the ``first'' focal level).
FSO have been applied, albeit in a limited and simplified formulation, in
the course of iMinds project ``Little Sister''~\cite{LS,DFSB13a}, and are
at the core of recently launched project ``SELFSERV''~\cite{Selfserv}.

FSO have been studied, to some preliminary extent, in~\cite{DF13c}.
In the cited paper I introduced a mathematical model, extending
results already obtained in other works~\cite{DeFl96,DeFl05}, in which
the evolution of the entities in an FSO hierarchy is represented as
a random walk among the set of all possible SONs resulting from
the collaboration of all FSO entities.

\begin{figure*}
	\centerline{\includegraphics[width=0.37\textwidth]{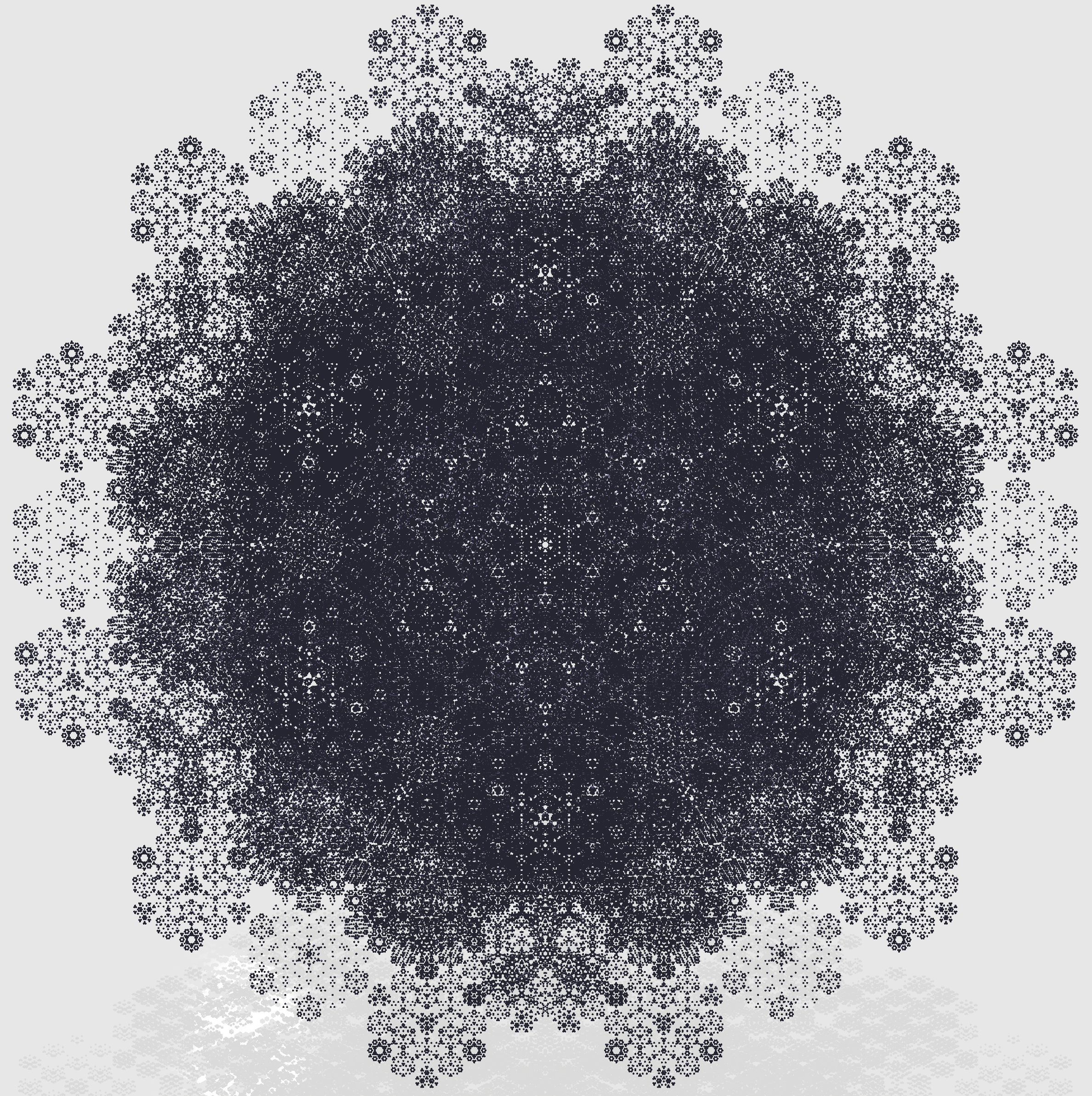}%
	            \includegraphics[width=0.37\textwidth]{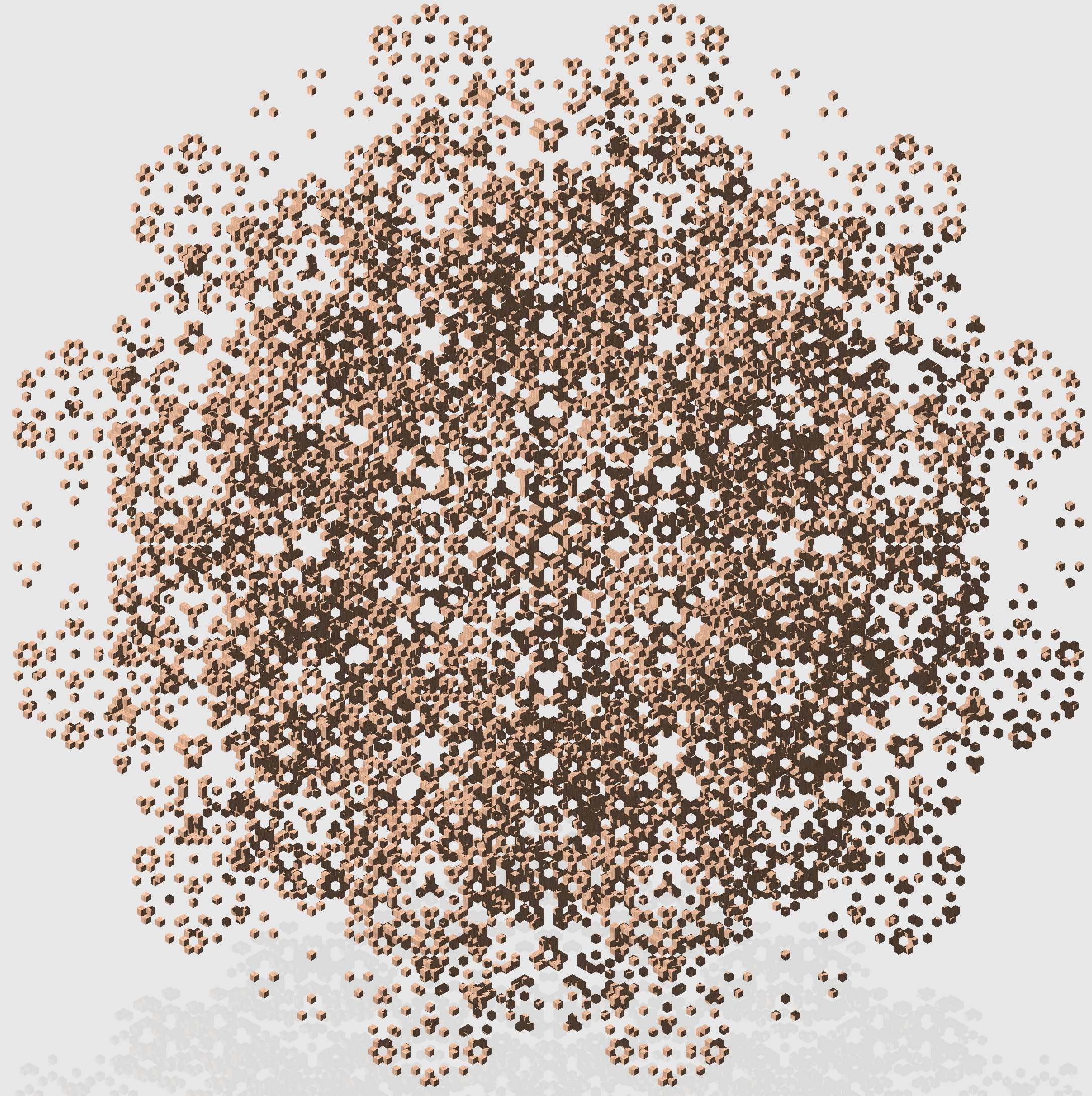}%
                   }
	\caption{Fractal social organizations 011112233334 (left-hand side)
and 011123334. The modularity emerging in the geometrical representations
is a direct consequence of the modularity in the ``seeds''.}
\label{f:fso}
\end{figure*}

I deem several facts to be relevant to the current discussion:
\begin{itemize}
\item The mentioned mathematical model takes as input a ``flat'' set of
entities, irrespective of their hierarchical position.
Despite this flat initial configuration, the space of all possible SONs produced by the model
is hierarchical; modular; self-similar; and admits a fractal dimension\cite{DF13c}.
Figure~\ref{f:fso} exemplifies two FSOs.

\item Representations of the space of all possible SONs may be considered
as phenotypical representations of a genotypical ``code'', or ``seed'',
given by a string identifying all the possible roles in the FSO.

\item A property of the FSO representations is given by
\emph{conservation of modularity}: if the FSO seed $a$ is a sub-string of FSO seed $b$, then
the FSO development of $b$ includes, and \emph{is a refinement of}, the FSO
development of $a$. Figure~\ref{f:fso} exemplifies this with
$b=011112233334$ and $a=011123334$.
\end{itemize}

Because of the above observations, 
I argue that pictures such as those in Fig.~\ref{f:fso}
provide a geometrical interpretation of the Leibnitian concept of monad
as well as an exemplification of the static 
hierarchy of the biota introduced by Hierarchy Theory.

\section{Elasticity, Resilience, and Antifragility}\label{s:res}
A third field worth of discussion in the framework of this paper is
resilience and its related methods.

As discussed in~\cite{DF14d},
resilience may be defined as ``a system's ability to either absorb or tolerate change without losing
one's peculiar traits or expected behaviors''. As observed in the cited reference,
this definition consists of
two parts corresponding to the following two features:
\textbf{Evolvability}, i.e.,
the ability to ``alter [one’s] structure or function so as to adapt to
changing circumstances''~\cite{Jen:2003}.
\textbf{Identity persistence}: an evolving system's ability to retain one's features and
characteristics in spite of exogenous and endogenous changes.
The above two conditions closely correspond to the Aristotelian concept of
entelechy. An entelechy is in fact
a subject that ``brings about their own changes from one state to another''
and, at the same time, one that
``exercises activity in order to guarantee one's identity''
or ``to comply to one's \emph{definition}''~\cite{DeAnimaLawsonTancred}.
The above two constituent aspects are elegantly rendered by Sachs'
translation of entelechy as
``being-at-work while staying-the-same''~\cite{Sachs}. 
This highlights 
the two constituent properties of a resilient system or being:
(1)
being able to persist one's uniqueness---one's identity---throughout time;
(2) being able to manifest/construct and persist oneself in the physical world.
We end up with another dualism, strictly related to the System of Leibniz 
discussed in Sect.~\ref{s:Leibniz}.


Now, manifesting/constructing one's identity requires a design; and due to economic
considerations related to physical and/or biological constraints, said design calls
for the adoption of trade-offs regarding the practical organization of a system or being.
(cf. for instance~\cite{Nilsson08} and~\cite{WeWi55} for practical examples of
and justifications for the introduction of these design trade-offs).
Thus ``materialization'' (in the sense elucidated in Sect.~\ref{s:Leibniz})
inherently implies a
hard coupling with a ``reference environment'', namely an hypothesized
set of average ``operational'' environmental conditions.
Resilience refers then to the following two major abilities:
(1) The (static) ability to absorb fluctuations in the experienced
environmental conditions. This ability is known as \emph{elasticity\/}
and corresponds to being able to mask change through the
adoption of a predefined
amount of redundant resources.
(2) The ability to adjust dynamically one's ``materialization'' and to operate
actions, both individually and socially, such that changes are tolerated.
I call this ability \emph{entelechism}.

From a systemic point of view, elasticity corresponds to simple purposeful behaviors,
namely behaviors that are non-teleological~\cite{RWB43}.
On the contrary, entelechism
mandates complex cybernetic behaviors embedding feedback loops to guide one's
action towards the intended goals~\cite{RWB43}.

The difference between the two approaches may be better understood when considering
a particular context. In the framework of information theory, for instance,
a well-known result
by Shannon~\cite{Shannon} tells us that it is possible to transfer reliably
a message across an unreliable channel by replicating ``sufficiently'' the
message. 
The problem is of course learning what is the ``right amount'' of
replicas. A strategy to deal with this problem is to monitor
the channel and identify a ``worst case'': we could find out for instance
that five replicas of each message are sufficient to guarantee reliable
transmission across the channel. This is a typical elastic strategy.
The major benefit of elasticity is simplicity: the transmitter does not
need to enact any complex behavior and it can be completely unaware
of the actual condition of the channel---in other words, no interaction
with the environment is either foreseen or required. 
Elasticity is pure and context-agnostic replication.
Two are the negative aspects
of elasticity:
First, it is based on a ``snapshot'' of the environment taken in the past.
In other words, it refers to a situation that possibly has changed.
Secondly, it equalizes all actual conditions to the worst case.
Thus if the worst case only occurs occasionally, elasticity makes
use of an unnecessarily high amount of resources.

From the above discussion I conjecture that the HT role of \textbf{replicator}
closely corresponds to elastic strategies.
This may become more apparent 
when considering the case of transmission
through a temporal (or better, genealogical) channel rather than through
a spatial channel.

A second strategy is to have the transmitter monitor the channel;
extrapolate the unreliability of the channel during the next transmission time; and
adopt a degree of redundancy best-matching the extrapolated conditions.
This complex teleological behavior is one of entelechism. 
Entelechism implies a complex interaction with the environment and the ability
to enact extrapolations. It also implies the ability to
enact corresponding measures to reach the intended goal---in this case,
reliable transmission. This extra complexity
is the major negative aspect of entelechism.

From the above discussion one can see that
entelechism is a strategy of open systems able to become aware of the 
environmental conditions and to exercise complex cybernetic behaviors on their
environment; because of this I draw an analogy between entelechies
and HT \textbf{Interactors}.

\subsection{Computational Antifragility}
As mentioned above, a key requirement of resilience is identity persistence.
In some cases, however, preserving the identity of the system does not
appear as the most desirable course of action. As the environmental conditions change,
it may make more sense to \emph{adjust\/} the identity of the system so as to adopt
a ``form'' more profitable with respect to the original identity.
Natural evolution provides a clear example of the benefits of such a strategy.
Recent works~\cite{Jones2014870} show a growing interest towards
a different approach to resilience based on system identity evolution.
In~\cite{DF14a} I discussed this problem and hypothesized that
a possible strategy should combine elasticity, entelechism, and (machine) learning.
Computational antifragility is the name I use
for this class of strategies~\cite{DF14a,Taleb12}.
Whatever the design direction, we need to look at nature and change
the paradigm of resilience from ``being at work while staying the same'' to
``being at work while getting better'', namely becoming a better system or being.
With Leibniz's words we could say that, through
``a certain divine mathematics''~\cite{Leilines} we need to learn how nature
makes it possible that ``something greater will emerge''~\cite{LeibnizParkinson}.

\section{Conclusions}\label{s:end}

%

Paraphrasing the Authors of~\cite{HT:TE14a}, I am convinced that
``Incorporating insights from the hierarchy theory of evolution and from network theory provides a more
complete theoretical framework for explaining complex patterns and processes of biological'' (and artificial) ``evolution''.
Quoting Kenneth Boulding, this paper contributes to this process by
``point[ing] out similarities in the theoretical constructions of different disciplines''
and in particular
providing preliminary elements for
discussion about the implications of Hierarchy Theory
on the design and evolution of artificial systems and socio-technical organizations.
The engineering of resilient communitarian responses to crises and disasters and, in general, the
design of more resilient and antifragile
socio-technical systems and organizations constitute a natural
direction for the profitable application of the insights gathered through this
and other cross-disciplinary discussions.
Recently started project ``SELFSERV'' provides another direction,
in which a fractal social organization for the optimal management of diabetes
is being proposed~\cite{Selfserv}.

\section*{Acknowledgements}
We gratefully acknowledge support from VLIR-UOS through Project SELFSERV.



%


\bibliographystyle{model3a-num-names}



\clearpage

\normalMode

\end{document}